\documentclass[fleqn,usenatbib]{mnras}
\usepackage{newtxtext,newtxmath}

\usepackage[T1]{fontenc}
\usepackage{ae,aecompl}
\usepackage{flushend}
\usepackage{xcolor}

\usepackage{graphicx}	
\usepackage{amsmath}	
\usepackage{amssymb}	
\newcommand{\fe}{f_{\rm esc}}

\title[Lyman Continuum Escape Fraction]{Lyman Continuum Escape Fraction in Ly$\alpha$ Emitters at $z\simeq3.1$}

\author[Fuyan Bian \& Xiaohui Fan]{
Fuyan Bian$^{1}$\thanks{E-mail: fbian@eso.org (FB)},
Xiaohui Fan$^{2}$
\\
$^{1}$European Southern Observatory, Alonso de C\'ordova 3107, Casilla 19001, Vitacura, Santiago 19, Chile\\
$^{2}$Steward Observatory, University of Arizona, 933 N Cherry Ave., Tucson, AZ 85721, USA
}

\date{Accepted XXX. Received YYY; in original form ZZZ}

\pubyear{2015}

\begin{document}
\label{firstpage}
\pagerange{\pageref{firstpage}--\pageref{lastpage}}
\maketitle

\begin{abstract}
We measure the LyC escape fraction in 54 faint Lyman Alpha Emitters (LAEs) at $z\simeq3.1$ in the GOODS-South field. With the average magnitude of $R=26.7$ AB ($M_{UV}=-18.8$, $L\simeq0.1L^*$), these galaxies represent a population of compact young dwarf galaxies. Their properties are likely to resemble those in the galaxies responsible for reionising the Universe at $z>6$. We do not detect LyC emission in any individual LAEs in the deep {\em HST} F336W images, which covers the rest-frame 820{\AA}. We do not detect the LyC emission of these LAEs in the stacked F336W images, either. The $3\sigma$ upper limit of LyC escape fractions is $\fe<14-32\%$. However, the high Ly$\alpha$ rest-frame equivalent width, low stellar mass and UV luminosity of these LAEs suggest that they should have $\fe>50\%$. The low LyC escape fraction from this work and other stacking analysis suggest that the LyC leaking galaxies with $\fe > 50\%$ at $z=2-3$ do not follow the relation between the $\fe$ and UV luminosity and Ly$\alpha$ equivalent width (EW) derived from typical galaxies at similar redshift. Therefore, the UV luminosity and Ly$\alpha$ equivalent width (EW) are not the best indicators for the LyC escape fraction.
\end{abstract}




\section{Introduction}
The epoch of reionisation is a period when neutral hydrogen in the intergalactic medium (IGM) was ionised by the first generation energetic sources in the Universe. Current observations have relatively well constrained the cosmic reionisation history, which occurred at the redshift of $z=7-10$ and finished largely by $z=6$ \citep[e.g,][]{Fan:2006lr,Fan:2006aa,Stark:2011fk,Schroeder:2013aa,Schenker:2014aa,Bian:2015aa,Bouwens:2015ad,Robertson:2013dq,Planck-Collaboration:2016aa,Banados:2018aa,Itoh:2018aa}. However, it is still under debate what are the major sources that reionise the Universe due to the following two main uncertainties: (1) the faint-end luminosity function of star forming galaxies \citep[e.g,][]{Atek:2018aa,Bouwens:2017ab} and active galactic nuclei (AGN) \citep[e.g.,][]{Giallongo:2015aa,McGreer:2018ab,Boutsia:2018aa,Matsuoka:2018aa} at high redshift, and (2) Lyman continuum (LyC) escape fraction ($\fe$) in galaxies, the fraction of the ionising photons (<912{\AA}) that can escape from a galaxy to reach the IGM. If star-forming galaxies are the major sources to reionise the Universe, it requires a LyC escape fraction at least $\fe=0.2$ at the epoch of reionisation \citep[e.g.,][]{Robertson:2013dq,Naidu:2019aa}, by adopting a typical IGM clumping factor \citep[e.g.,][]{Pawlik:2009aa}, galaxy luminosity function at $z=7$ \citep[e.g.,][]{Atek:2015ab,Atek:2018aa,Bouwens:2015ab,Bouwens:2017ab,Livermore:2017aa}, and LyC photon production efficiency \citep[e.g.,][]{Bouwens:2016ab,Tang:2018aa,Chevallard:2018aa}.

However, LyC escape fraction can not be directly measured in galaxies beyond $z=4.5$ due to the high opacity of the IGM to LyC ionising photons \citep[e.g.,][]{Vanzella:2018aa}. Thus we have to infer the LyC escape in the galaxies at the epoch of the reioisation based on either directly measuring LyC escape fraction in galaxies at lower redshift or correlating galaxy spatial positions with the Lyman alpha forest at $z>6$ \citep[e.g.,][]{Kakiichi:2018aa}. In the last decade, people have conducted extensive studies of LyC escape fraction in galaxies at $z<4$ using a number of different approaches, including the rest-frame ultraviolet spectroscopy \citep[e.g.,][]{Leitet:2013aa,Leitherer:2016aa,Izotov:2016aa,Izotov:2016ab,Steidel:2001aa,Steidel:2018aa,Shapley:2006kx,Shapley:2016aa,Nestor:2013aa} and  narrow/intermediate/broad-band UV imaging \citep[e.g.,][]{Siana:2007aa,Siana:2015aa,Vanzella:2010aa,Vanzella:2016ab,Nestor:2011ab,Cooke:2014aa,Rutkowski:2016aa,Rutkowski:2017aa,Vasei:2016aa, Bian:2017ab,Japelj:2017aa,Naidu:2017aa,Fletcher:2018aa,Ji:2019aa}. 
However, accurately measuring the escape fraction remains difficult. Most of these studies yielded null or tentative detection of LyC emission. Furthermore, studies based on ground-based observations suffer from foreground contamination, resulting in overestimating the LyC escape fraction. To date, there are only several convincing detection of LyC emission in galaxies at $z\sim3$, including $Ion2$ and $Ion3$ \citep{Vanzella:2016ab,Vanzella:2018aa}, Q1549-C25 \citep{Shapley:2016aa}, A2218-Flanking \citep{Bian:2017ab}, and Sunburst Arc \citep{Rivera-Thorsen:2019aa}. In addition, the LyC escape fraction measured in individual galaxies has large uncertainty due to the opacity variations of the line of sight of IGM and galaxy ISM \citep{Cen:2015aa}. Furthermore, the high LyC escape fraction measured in individual objects is not necessarily representing the LyC escape fraction of typical galaxies at the epoch of reionisaition.

Studies have shown that the high-redshift galaxy luminosity function is steep at the faint end, thus sub-$L^*$ galaxies dominate the UV emission at the epoch of reionisation. It is essential to push the LyC escape fraction measurement to faint galaxies. In this study, we measure the LyC escape fraction in a sample of Ly$\alpha$ emitters (LAEs) at $z\simeq3.1$ with $L\sim0.1L^*$ in the GOODS-South field \citep{Dickinson:2003aa}. The redshift of these LAEs have been accurately measured by the MUSE Hubble Ultra-Deep Field (HUDF) survey and the MUSE-Wide survey based on their Ly$\alpha$ emission lines \citep{Japelj:2017aa,Urrutia:2018aa}. Their LyC emission is well covered by the deep HST/WFC3 F336W images from the Hubble Deep UV Legacy Survey \citep[HDUV,][]{Oesch:2018aa}.

Throughout this paper, we use the following cosmological parameters: Hubble constant $H_0=70$~km~s$^{-1}$~Mpc$^{-1}$, dark matter density $\Omega_{\rm M}=0.30$, and dark energy density $\Omega_\Lambda=0.70$ for a flat universe. All the magnitudes are expressed in the AB magnitude system \citep{Oke:1983qy}.

\section{Sample Selection}
In this study, we use the deep {\em HST}/WFC3 F336W images to measure the LyC escape fraction in a sample of LAEs at $z\simeq3.1$ in the GOODS-South field. We use the HST/WFC3 F336W imaging data from the HDUV survey \cite[][]{Oesch:2018aa}. The HDUV survey is a deep UV imaging legacy survey covering a total of area of about 100~armin$^2$ in the two GOODS fields in F275W and F336W bands. The depths of the F275W and F336W bands are 27.6 and 28.0 (5$\sigma$), respectively. The deep F275W and F336W images from the HDUV survey have been used to study the LyC escape fraction in galaxies at $z\sim2$ and $z\sim3$, respectively \citep{Naidu:2017aa, Rutkowski:2017aa,Japelj:2017aa}. In particular, \citet{Japelj:2017aa} studied the LyC escape fraction in a sample of galaxies at $z=3-4$ selected in 9 arcmin$^2$ MUSE HUDF survey field using the HDUV F336W image and found the relative escape fraction $f_{\rm esc,rel}$ < 0.6 for galaxies with $0.1L^*$. Here we extend the study to a 44 arcmin$^2$ area using the newly published the MUSE wide survey \citep{Urrutia:2018aa}.
The MUSE-Wide survey is an integral filed spectroscopic survey, largely overlaps with the HDUV in the GOODS-South field. Thanks to the VLT/MUSE integral field spectrograph \citep{2010SPIE.7735E..08B}, this survey is able to detect galaxies without photometric pre-selection, thus it provides a high complete rate on detecting emission line galaxies. In the MUSE-Wide survey, we select LAEs in the redshift range of $z= 3.02-3.24$. At this redshift range, the HST WFC3/F336W filter cover the Lyman continuum at the wavelength range of 750 to 890{\AA}. In this study, we only use galaxies with redshift quality greater on equal to 2 \citep{Urrutia:2018aa}. The lower limit redshift of $z=3.02$ corresponds to the Lyman limits at observer-frame 3684 {\AA}. At this wavelength the throughput of the F336W filter is less than 1\%, which minimises the contamination of the flux redward of the Lyman limit.

A total of 54 Ly$\alpha$ emitting galaxies (LAEs) at the redshift range of $z= 3.02-3.24$ are selected in the GOODS-south field with deep HDUV F336W images coverage. Among them, 26 LAEs are selected from the MUSE wide survey \citep{Urrutia:2018aa}., and 28 LAEs are from the MUSE HUDF survey \citep{Japelj:2017aa}.
Their median stellar mass is $\log(M_*/M_{\sun})=8.0$, and median dust extinction is $A_V=0.1$. These LAEs represent a population of the galaxies with low stellar mass and young stellar population \citet{Urrutia:2018aa}. 

\begin{figure}
    \centering
    \vspace*{-.15in}
    \includegraphics[width=0.48\textwidth]{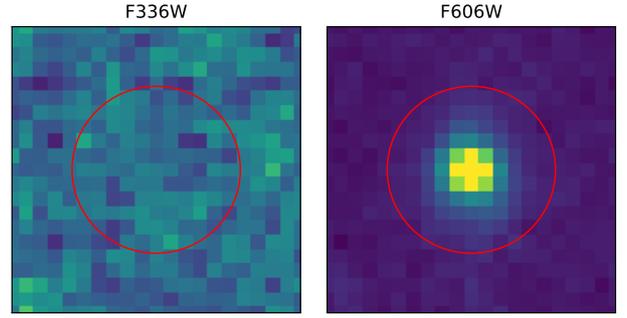}
    \vspace*{-.5in}
    \caption{Stacked images for the 54 LAEs at $z=3.1$ selected in the GOODS-S filed in {\em HST}/WFC3 F336W band (left) and ACS F606W band (right). The flux in each image is measured in the red circles with $0.7''$ in diameter. }
    \label{fig:image}
\end{figure}

\section{The Lyman Continuum Escape Fraction in LAEs at Z=3.1}

The {\em HST}/WFC3 F336W image covers the LyC emission at the rest-frame wavelength of $\sim820${\AA} in the LAEs at $z\simeq3.1$. None of the 54 Ly$\alpha$ emitting galaxies (LAEs) is detected in the F336W image at 3$\sigma$.
Therefore, we try to detect and measure the LyC emission by stacking the F336W images. We generate a $10''\times10''$ stamp image in F336W for each individual LAE based on the galaxy coordinate measured in the F606W images. Here we do not use the galaxy coordinate from the MUSE-Wide survey, because there exist small offsets ($\sim0.3''$) between the MUSE coordinates and the HST coordinates \citep{Urrutia:2018aa}. The stacked F336W image is generated by combining the F336W stamp images of all 54 LAEs using the mean flux at each pixel (Figure~\ref{fig:image}). We measure the LyC flux in the F336W stacked image using a $0.7''$ diameter aperture and find that LyC flux is not detected at 3$\sigma$ level. The $3\sigma$ flux upper limit in F336W band is $<0.002\mu$Jy, which corresponds to $>30.64$ AB magnitude. We use the flux in the HST/ACS F606W image, which corresponds to the rest-frame wavelength of 1500{\AA}, as an anchor point to estimate the intrinsic LyC emission at 820{\AA}. We use the ACS F606W images from the 3D-HST data release 4.5.1\footnote{https://3dhst.research.yale.edu/Data.php} \citep{Momcheva:2016aa}. We combine the F606W images for the LAEs following the same procedure used for the F336W images (Figure~\ref{fig:image}). Then the F606W flux is measured in the final combined F606W images using a $0.7''$ diameter aperture. This aperture size is the same as used in the 3D-HST survey \citep{Skelton:2014aa}, which includes a large fraction of the flux from the object and avoid flux from the neighbouring objects. The flux density of the F606W band is $0.074\pm0.001\mu$Jy. The apparent magnitude in F606W band is $26.72\pm0.01$ AB magnitude, and the absolute magnitude at the rest-frame wavelength of 1500{\AA} is $M_{UV}=-18.86$, which corresponds to $0.1L^*$ at $z\sim3$ \citep[e.g.,][]{Bian:2013aa}. 

We measure the relative LyC escape fraction $f_{\rm esc}$:
\begin{equation}
    f_{\rm esc, rel} = \frac{L_{1500}/L_{820}}{f_{1500}/f_{820}}\times \exp(\tau_{\rm IGM,820})
\end{equation}
and absolute LyC escape fraction
\begin{equation}
    f_{\rm esc} = f_{\rm esc, rel} \times 10^{-0.4A_{1500}}
\end{equation}
where $f_{1500}/f_{820}$ is the observed flux density ratio between 1500{\AA} and 820{\AA} measured in the HST F606W and F336W images. $L_{1500}/L_{820}$ is the intrinsic luminosity density ratio between 1500{\AA} and 820{\AA}. Here, we adopt two $L_{1500}/L_{820}$ values, $L_{1500}/L_{820} = 3$ and $L_{1500}/L_{820} = 7$, to cover a wide range of possible star formation history \citep[e.g.,][]{Siana:2007aa,Chisholm:2019aa}. $\exp(\tau_{\rm IGM,820}) \equiv 1/T$ is the mean IGM opacity at 820{\AA} at $z=3.1$ \citep[e.g.,][]{Madau:1995lr}, and $T$ is the mean IGM transmission. In this study, we adopt the \citet{Inoue:2014aa} recipe to estimate the mean IGM opacity and find $\exp(\tau_{\rm IGM,820})=3.84$ at $z=3.1$, corresponding to $T=0.26$. $A_{1500}$ represents the total dust extinction at 1500{\AA}, which is estimated from the median extinction at the V-band, $A_v=0.18$, by adopting the \citet{Gordon:2003aa} Small Magellanic Cloud (SMC) dust extinction curve. At last, we find that the 3$\sigma$ upper limit of average LyC escape fractions in this sample of LAEs are $\fe<14\%$ and $\fe<32\%$, for $L_{1500}/L_{820} = 3$ and $L_{1500}/L_{820} = 7$, respectively.

\begin{figure}
    \centering
    \includegraphics[width=0.47\textwidth]{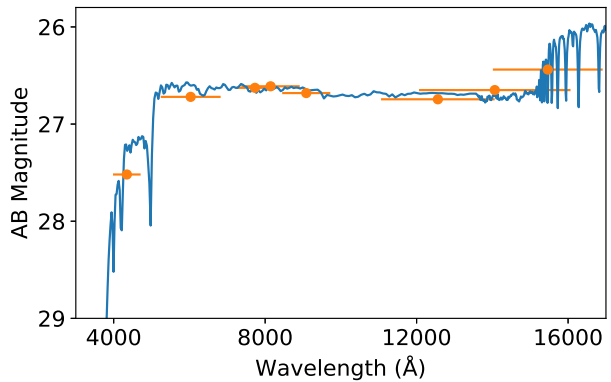}
    \vspace*{-.15in}
    \caption{Best fit stellar synthesis model (blue curve) to the mean SED of the LAEs in this work in F435W, F606W, F775W, F814W, F850LP, F120W, F140W, and F160W bands (orange data points).}
    \label{fig:sed}
\end{figure}

\begin{figure}
    \centering
    \vspace*{-.2in}
    \includegraphics[width=0.47\textwidth]{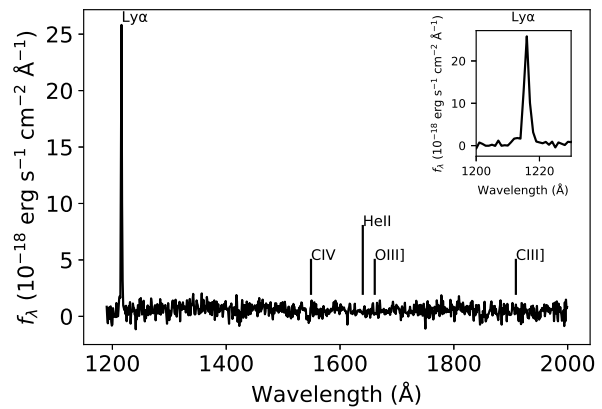}
    \caption{Composite spectrum of the LAEs whose LyC escape fraction is measured in the GOODS-South field. The strong emission line is the Ly$\alpha$ emission line.}
    \label{fig:spec}
\end{figure}

\section{Physical Properties of the Lyman-alpha Emitters}
We study the physical properties of the LAEs in this work by fitting their mean spectral energy distribution (SED). We obtain the mean SED by stacking the HST F435W, F606W, F775W, F814W, F850LP, F120W, F140W, and F160W images. All the images are taken from the 3D-HST data release 4.5.1 \citep{Momcheva:2016aa}. The images from each filter are combined by following the same method that is used to combine the F336W band images. The fluxes in each of these bands are measured within an aperture of $0.7''$ in diameter. The Fitting and Assessment of Synthetic Templates (FAST) code is used to fit the mean SED \citep{Kriek:2009fk}. We adopt the \citet{Bruzual:2003aa} stellar synthesis models with a Chabrier Initial Mass Function \citep[IMF;][]{Chabrier:2003wd}, an SMC dust extinction curve \citep{Gordon:2003aa}, a 0.2 solar metallicity (Z=0.004), and an exponentially declining star formation history. Figure~\ref{fig:sed} shows the best fit stellar synthesis model to the mean SED. We find the stellar mass of $\log(M/M_{\sun})=8.6$, the galaxy age of 120 Myr, and the average SFR of 4 $M_{\sun}$~yr$^{-1}$. The average SFR and stellar mass derived from the SED fitting place these galaxies on the galaxy star formation main sequence at $z\sim3$ \citep{Speagle:2014aa}.

The morphology of the LAEs is compact in the stacked F606W image (Figure~\ref{fig:image}), corresponding to the rest-frame UV ($\sim1500$\AA) images. We measure the galaxy size by fitting the stacked F606W image to a S\'ersic profile using the {\sc GALFIT} code \citep{Peng:2010bh}. The angular size of the galaxy effective radius is $r_e=0.16''$, corresponding to physical size of $r_e=1.2$~kpc. The S\'ersic index derived from the fitting is $n=1.37$, suggesting an exponential disk-like morphology. The mean galaxy effective radius and mass place these galaxies on the mass-size relation at $z\sim3$ when extrapolating the relation to low mass end \citep[e.g.,][]{van-der-Wel:2014aa}.

We obtain a composite spectrum of the LAEs by stacking the MUSE spectra. First, we de-redshift each MUSE spectrum to the rest-frame wavelength and re-sample spectrum to rest-frame wavelength from 1180{\AA} to 1950{\AA} with an interval of 1{\AA}. The spectra were combined using the average flux density at each wavelength. Figure~\ref{fig:spec} shows the composite spectrum of the LAEs. The rest-frame equivalent width of Ly$\alpha$ emission is EW$_0 = 142${\AA}. The high Ly$\alpha$ EW$_0$ suggests that the ages of these LAEs are young, on the order of a few times $10^7$ years for a constant SFR model \citep{Hashimoto:2017aa}. This is broadly consistent with the SED fitting results. The Ly$\alpha$ flux is $5.1\times10^{-17}$~erg~s$^{-1}$~cm$^{-2}$, corresponding to a luminosity of $4.2\times10^{42}$~erg~s$^{-1}$. The the Ly$\alpha$ emission escape fraction is $\sim60\%$ based on the SFR of 4~$M_{\sun}$~yr$^{-1}$ \citep{Kennicutt:1998aa, Verhamme:2017aa}.

\begin{figure}
    \centering
    \includegraphics[width=0.47\textwidth]{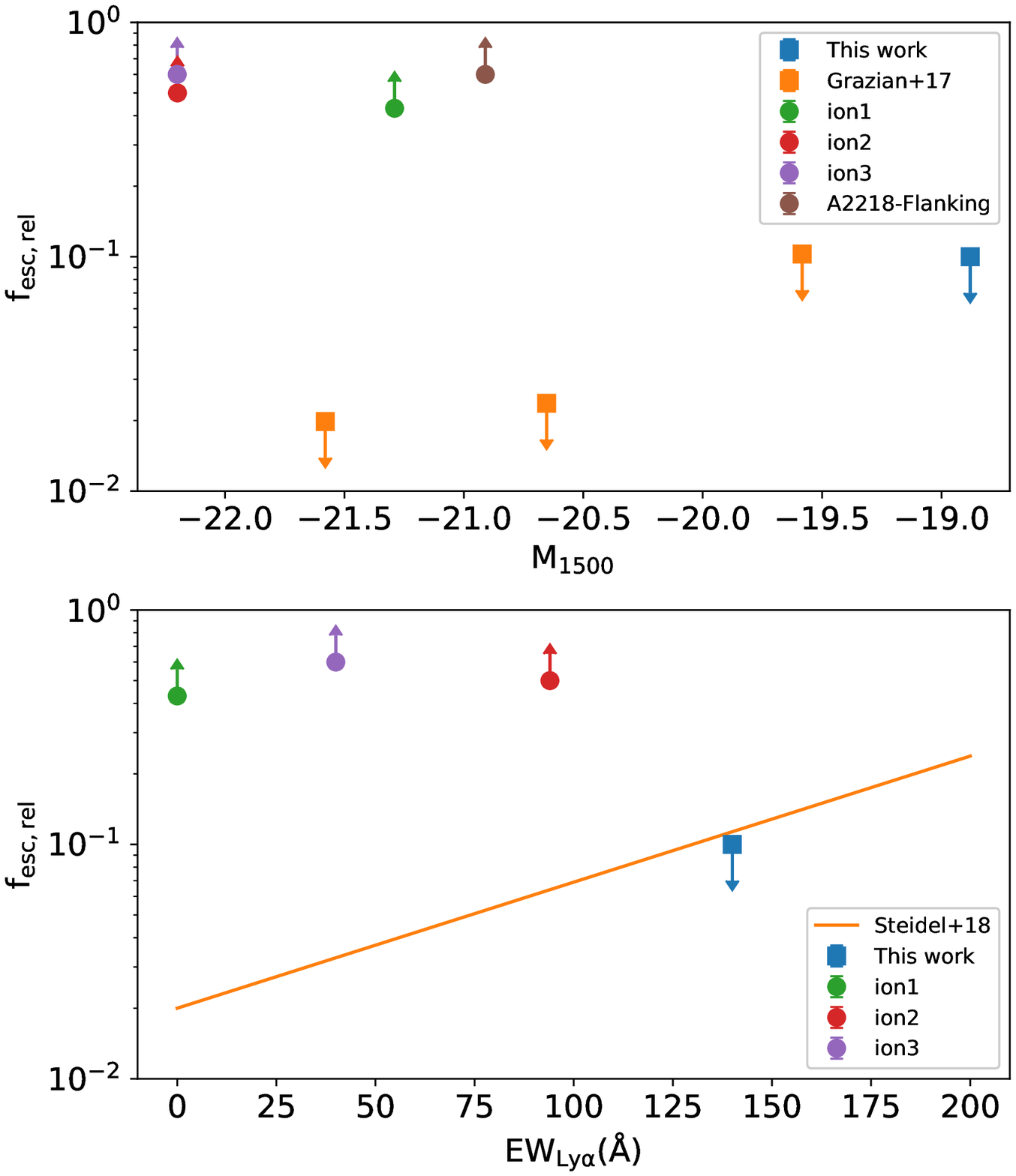}
    \vspace*{-.1in}
    \caption{Relative LyC escape fraction vs absolute UV magnitude (upper panel) and EW of Ly$\alpha$ line (bottom panel). The squares represent the measurements from stacked images \citet{Grazian:2017aa} and this work, and the the solid circles represent the results from individual galaxies, {\it Ion1} \citep{Ji:2019aa}, {\it Ion2} \citep{Vanzella:2016ab}, {\it Ion3} \citep{Vanzella:2018aa}, and A2218-Flankin \citep{Bian:2017ab}. The orange solid line is the relation adopted from \citet{Steidel:2018aa} based on the stacked spectra. All the relative LyC escape fraction has been scaled to $L_{\rm UV}/L_{\rm LyC}=3$ and 1$\sigma$ the upper limit.}
    \label{fig:fesc_prop}
\end{figure} 

\section{Discussion}

\subsection{LyC emission and Galaxy Properties}
The relations between the LyC escape fraction and galaxy properties have been established by both theoretical and observational studies\citep[e.g.,][]{Reddy:2016aa,Verhamme:2017aa,Steidel:2018aa}.
Galaxies with stronger Ly$\alpha$ emission strength, weaker interstellar absorption strength, higher [OIII]/[OII] ratio, higher star formation surface density, lower stellar mass and UV luminosity tend to have higher escape fraction \citep[cf.][]{Bassett:2019aa,Ji:2019aa}. Such studies help us to better understand the physical mechanism(s) that drives the LyC leaking from a galaxy; extrapolation of such relations to galaxies at higher redshift can be used to predict the LyC escape in galaxies at the epoch of reionisation.

In this study, the galaxies are selected by their strong Ly$\alpha$ emission lines. These LAEs show high Ly$\alpha$ equivalent width and high Ly$\alpha$ escape fraction. Studies have suggested that galaxies with high LyC escape fractions commonly shows strong Ly$\alpha$ emission (EW$_0$(Ly$\alpha) > 70${\AA}) and high Ly$\alpha$ emission line escape fraction $>20\%$ \citep[e.g.,][]{Verhamme:2017aa,Kimm:2019aa}. To our knowledge, the Ly$\alpha$ EW$_0$ in the composite spectrum of these LAEs is higher than all of the known LyC-leaking galaxies with $\fe>30\%$ \citep[e.g.,][]{Verhamme:2017aa,Vanzella:2016ab,Vanzella:2018aa}(Figure~\ref{fig:fesc_prop}). The other properties of the LAEs in this work are also in favour to having high LyC escape fraction, including lower stellar mass and UV luminosity, compared to those individual galaxies with high LyC escape fraction ($\fe>30\%$) at the similar redshift range, such as {\it Ion2}, {\it Ion3}, and A2218-Flank (Figure~\ref{fig:fesc_prop}). Based on these properties, these LAEs in this study are expected to have even larger LyC escape fraction $>30\%$. However, the non-detection of LyC emission in the stacked HST F336W image suggests a rather low ($<14-32\%$) LyC escape fraction in these LAEs. 

Figure~\ref{fig:fesc_prop} shows the relative LyC escape fraction vs. absolution UV magnitude and Ly$\alpha$ EW. The relative LyC escape fraction measured in the individual galaxies have much higher LyC escape fraction measured in the stacked images or spectra. The individual galaxies with high LyC escape fraction do not follow the relation of the LyC escape fraction and the galaxies physical properties derived from the stacking analysis based on typical $z\sim3$ galaxies. Therefore, their high escape fractions may not represent the typical LyC escape fraction in the galaxies with similar UV luminosity and Ly$\alpha$ EW at $z\sim3$. It suggests that the UV luminosity and Ly$\alpha$ EW may not be the best indicator for the LyC escape fraction, especially for galaxies with high LyC escape fractionn. Actually, it has been found that in some of the cases the Ly$\alpha$ EW itself is not a good indicator of LyC escape fraction \citep[e.g.,][]{Guaita:2016aa,Grazian:2017aa,Ji:2019aa}, because the Ly$\alpha$ photons can escape via pure radiative transfer effects even in a relatively high \ion{H}{I} column density (e.g., $N_{\rm HI}>10^{20}$~cm$^{-2}$), which is completely optically thick for the LyC radiation \citep[e.g.,][]{Verhamme:2006fk}. Therefore, the detailed Ly$\alpha$ profile, such as the separation of the blue and red peaks and leaking of the Ly$\alpha$ at systematic redshift, can also provide crucial information on the LyC leaking.


\subsection{Systematic Uncertainties of LyC Escape Fraction}
Our LyC escape fraction measurement also faces large systematic uncertainties as follows:

1. Galaxy viewing angle:  Studies have shown that the LyC photons can only escape from chimneys and holes in the ISM of galaxy caved by the supernovae and other stellar feedbacks \citep[e.g.,][]{Heckman:2011aa}. This indicates that the leaking LyC photons can only be detected in a small fraction of the solid angle of galaxies, and the LyC escape fraction measurements highly depends on the viewing angle of the galaxy. It can cause a large uncertainty on the LyC escape fraction based on the measurements in a small sample of galaxies. \citet{Cen:2015aa} suggested that it requires to stack at least 100 galaxies to reduce the LyC escape fraction uncertainty down to 20\%. 

2. IGM transmission: 
LyC escape fraction measurement also depends on the IGM transmission. \citet{Japelj:2017aa} found that the $z=3.1$ IGM transmission in F336W band at a random line of sight is $0.26^{+0.30}_{-0.25}$. For a sample of 54 galaxies, the uncertainty of the IGM transmission is about 15\%.

3. The spatial offset between the LyC and UV light centriods: The stacking strategy relies on the assumption that the LyC emission has the same location as the UV emission at {1500\AA}. 
Observations of gravitational lensed system, Sunburst, indicate that the LyC emission only emerges in some of the star-forming knots and varies significantly from one knot to other knots \citep{Vanzella:2019aa,Rivera-Thorsen:2019aa}. Therefore, the LyC and UV centroid positions are not necessary always well-aligned. The LyC signal will be diluted due to such misalignment during the stacking process, and the LyC escape fraction can be significantly underestimated.


\section{Conclusions}
In this letter, we study the LyC escape fraction in a sample of LAEs. These LAEs represent a population of compact young dwarfs at $z=3.1$. We summarise the main results of this work as follows:

1. A sample of 54 Ly$\alpha$ emitters (LAEs) at $z=3.02 - 3.24$ are selected from the MUSE Hubble Ultra Deep Field (HUDF) and MUSE MUSE-Wide integral field spectroscopic survey in the GOODS-South field based on their prominent Ly$\alpha$ emission. 

2. We fit stellar synthesis models to the composite SED of these LAEs and find the stellar mass of $\log(M/M_{\sun})=8.6$, the age of 120 Myr, and the SFR of 3.2 $M_{\sun}$~yr$^{-1}$.

3. The galaxy size at the UV wavelength measured in the stacked HST F606W image is 1.2~kpc, and the rest-frame equivalent width of the Ly$\alpha$ emission is 142{\AA}, which is measured in the composite MUSE spectra of these LAEs.

4. The LyC emission of these LAEs is not detected at 3$\sigma$ level in the stacked HDUV deep HST F336W image, covering the rest-frame wavelength of 820{\AA} in these LAES. The upper limits of LyC escape fraction are $\fe<14\%$ and $\fe<32\%$, for $L_{1500}/L_{820} = 3$ and $L_{1500}/L_{820} = 7$, respectively.
    
5. Such low LyC escape fraction of these LAEs suggests that the LyC leaking galaxies at $z\sim3$ do not follow the relation of LyC escape fraction and galaxy properties, including UV luminosity and Ly$\alpha$ EW. It implies that the UV luminosity and Ly$\alpha$ EW are not the best properties to predict the LyC escape fraction, particularly for galaxies with high escape fractions.

\section*{Acknowledgements}
We thank the MUSE GTO, HDUV, and 3D-HST teams to release their data set to public, making this work possible. We thank Dr. J. Japelj to share the detailed information on the LAEs in the MUSE Hubble Ultra Deep Field Survey. We thank the anonymous referee for providing constructive comments and help in improving the manuscript. 




\bibliography{paper}




\bsp	
\label{lastpage}
\end{document}